\documentstyle[preprint,aps,prl,epsf]{revtex}
\begin{document}
\title{$J/\Psi$ Suppression in Pb--Pb Collisions: 
A Hint of Quark--Gluon Plasma Production?}

\author{Jean-Paul
Blaizot$^*$ and Jean-Yves Ollitrault\footnote{Affiliated to CNRS}}
\address{Service
de Physique Th\'eorique\footnote{Laboratoire de la Direction des Sciences
de
la Mati\`ere du Commissariat \`a l'Energie Atomique}, CE-Saclay \\ 91191
Gif-sur-Yvette cedex, France}
\maketitle

\begin{abstract}
The NA50 Collaboration has recently observed a strong 
 suppression of $J/\Psi$ production in Pb--Pb collisions at 158 Gev/n.
We show that this recent observation finds a quantitative explanation 
in a model which relates the suppression mechanism to
the local energy density, whose value is higher in  Pb--Pb collisions than in
any other system studied previously.  The
sensitivity of the phenomenon to small changes in the energy density 
could be suggestive of quark-gluon plasma formation.
\end{abstract}


The NA50 Collaboration has recently reported the observation of a strong
suppression of
$J/\Psi$ production in Pb--Pb collisions at 158 GeV
per nucleon \cite{NA50Mor96}, which is not explained by conventional models
of nuclear
absorption. Since such models have been found to account reasonably well 
for all the  previous data involving lighter nuclei\cite{GH92}, the immediate
implication seems to be that  new physics is involved in Pb--Pb collisions, 
possibly the formation of a  quark-gluon plasma \cite{Matsui86}.

In this note, we present an interpretation of the data based on the observation
that the {\it local energy  density\/} is higher  in Pb--Pb collisions than in
any of the systems studied previously, in particular the S--U system. In
order to explore the possibility that  the
large suppression which is observed in Pb--Pb collisions could be due to
the formation of a  quark-gluon plasma, we adopt a simplified description
\cite{BO88}, in which one considers that {\it all the
$J/\Psi$'s produced in a region where the energy density
 exceeds some critical value are suppressed\/}.  We
emphasize that we do not, and cannot at this stage, make precise statements
about detailed microscopic mechanisms (for a recent review, see \cite{KS95}).  
Our purpose is only to test the idea that
the suppression
depends solely on   the  local energy density, and to see whether
consequences of this assumption
are supported by the data.

We first review briefly the conventional treatment of
nuclear absorption \cite{Capella88,GH92}. The ratio of the
$J/\Psi$ production cross section in proton-nucleus collisions ($\sigma_{pA}$)
 to that in
proton-proton collisions ($\sigma_{pp}$) is given by
\begin{equation}\label{sigmapA}
{\cal
N}_A=\frac{1}{A}\frac{\sigma_{pA}}{\sigma_{pp}}=\frac{1}{A\sigma_a}\int{\rm
d}^2b\left(   1-\exp\left( -\sigma_a T_A({\bf b})\right)\right),
\end{equation}
where
$
T_A({\bf s})=\int_{-\infty}^{+\infty}\rho_A({\bf s},z) {\rm d}z
$
is the nucleon density per unit area in the transverse plane (i.e. the
plane transverse
to the collision axis), and $\sigma_a$ is an absorption cross section
(although we do not write it
explicitly, in our calculations $\sigma_a$ is multiplied
 by the correction factor $(1-1/A)$,
where $A$ is the mass number of the nucleus \cite{GH92}). The quantity
${\cal N}_A$ may be interpreted as the probability that a produced $J/\Psi$
survives nuclear absorption.

In a nucleus-nucleus collision, the survival probability ${\cal N}_{AB}$
 takes,  after integration over the impact parameter, the
 factorized form
 ${\cal N}_{AB}={\cal N}_A{\cal N}_B$. 
Within the range of $A$ values considered, and for the chosen
parametrization of the density,  
${\rm ln}{\cal N}_{AB}\sim A^{1/3}+B^{1/3}$ (see Fig.1).
We use for $\rho({\bf r})$ the expression:
$
\rho_A({\bf r})/\rho_0=1/\left(1+\exp\left(\frac{r-R_A}{a}\right)\right)
$
with $R_A=1.1 A^{1/3}$ fm,  $a=0.53$~fm, and $\rho_0$ is fixed by
the  normalization $\int\rho_A({\bf r})d^3r=A$ (e.g. in $^{208}$Pb,
$\rho_0=0.17$fm$^{-3}$).

As seen on Fig.1, nuclear absorption explains both the
proton-nucleus and the nucleus-nucleus data up to the S--U system, with a
common value of the absorption cross section (we have adopted the value
$\sigma_a\approx 6.2$~mb used by the NA50 Collaboration \cite{NA50Mor96}).
However, the  Pb--Pb system deviates significantly  from
that common trend. One can measure this deviation by the ratio
\begin{equation}\label{rpsi}
r_\Psi=\frac{{\cal N}_{AB}({\rm measured})}{{\cal N}_{AB}({\rm estimated})}\, ,
\end{equation}
where ${\cal N}_{AB}({\rm estimated})$ is the value of  the survival
probability to
nuclear absorption alone (${\cal N}_{AB}({\rm estimated})\approx 0.43$).
The value of $r_\Psi$, as read from Fig.1, is $0.68\pm 0.06$. 
The survival probabilities plotted in Fig.1 are extracted from 
absolute cross sections which contain systematic errors
of the order of 10 to 20\%. These errors have been reported
in the figure although they largely cancel in the relative values of 
the survival probabilities corresponding to a given set of data. 
In its analysis, the NA50 Collaboration uses the ratio of the $J/\Psi$
production cross section to the Drell--Yan cross section, which reduces most 
of these systematic errors. It also relies, in extracting nuclear
absorption, on the transverse energy dependence of the S--U data, rather 
than on integrated data alone. It obtains thus, for the ratio $r_\Psi$, the 
more precise value $0,72\pm 0.03$. 

We turn now to the dependence of the effect on  the impact parameter of the
collision. We write the $J/\Psi$
production cross section at impact parameter $b$ as follows:
\begin{equation}\label{sigmaABdeb}
\frac{1}{\sigma_{pp}}\frac{{\rm d}\sigma_{AB}}{{\rm d}^2b}=
T_{AB}(b){\cal N}(b).
\end{equation}
where $T_{AB}(b)=\int {\rm d}^2s\, T_A({\bf s})\,T_B({\bf s-b})$ is 
proportional to the probability to produce a $c\bar c$ pair, and 
${\cal N}(b)$ is the survival probability
at impact parameter $b$. The quantity ${\cal N}(b)$ is related to ${\cal
N}_{AB}$ introduced
above by
$
{\cal N}_{AB}=(1/AB)\int {\rm d}^2b\,T_{AB}(b){\cal N}(b).
$
If nuclear absorption is the only suppression mechanism,
\begin{equation}\label{Ndeb}
{\cal N}(b) ={\cal N}_{abs}(b)\equiv\frac{1}{T_{AB}(b)}\int {\rm
d}^2s\frac{1}{\sigma_a^2}
\left( 1-{\rm e}^{-\sigma_aT_A({\bf s})} \right)
\left( 1-{\rm e}^{-\sigma_aT_B({\bf s-b})} \right).
\end{equation}
In analogy with eq.(\ref{rpsi}), we define:
\begin{equation}\label{rpsideb}
r_\Psi(b)=\frac{{\cal N}(b)}{{\cal N}_{abs}(b)}.
\end{equation}
From the  NA50 data, assuming that the last bin in transverse energy
corresponds to central collisions, i.e. to $b\approx 0$, one extracts the value
$r_\Psi(b=0)\approx 0.50$.

We now show that the values of these two ratios, $r_\Psi$ defined in
eq.(\ref{rpsi}), and $r_\Psi(0)$ defined in eq.(\ref{rpsideb}), can be
understood quantitatively if one assumes that the suppression mechanism is
sensitive only to the local energy density. 
A central assumption here is that   the suppression of the
$J/\Psi$ takes place at  times short compared with the transverse size of the
interaction region, i.e.   before a substantial transverse expansion of the
produced matter has occurred, and before the
$J/\Psi$ has traveled a long distance in the transverse direction (we
consider  the production of
$J/\Psi$'s near central rapidity). We shall therefore
ignore both the transverse expansion and the transverse motion of
the $J/\Psi$'s. Under these conditions, the fate of a $J/\Psi$ 
is determined by the properties of the medium in the region where it is 
created, and is  controlled by the energy density in the transverse plane.

To estimate this density, we assume that it is proportional
to the density of participants. This assumption is motivated by the fact 
that in nucleus-nucleus collisions, the multiplicity and the transverse 
energy grow approximately linearly with the number of 
participants \cite{NA35}. The participants are the nucleons
which collide at least once  during the collision of nucleus A on nucleus B
at impact
parameter {\bf b}. They  have a density per unit transverse  area given by
\begin{equation}\label{npdes}
n_p({\bf s,b})=T_A({\bf s})\left[1-\exp\left(-\sigma_{N} T_B({\bf s-b})
\right)\right]
+T_B({\bf s-b})\left[1-\exp\left(1-\sigma_{N} T_A({\bf s})
\right)\right],
\end{equation}
where $\sigma_{N}\approx 32$ mb is the nucleon-nucleon inelastic cross section.
The total number of participants at impact parameter $b$ is 
$N_p(b)=\int{\rm d}^2s\, n_p({\bf s,b})$.

A plot of $n_p$ for the two systems S--U and Pb--Pb is given in
Fig. 2. One sees that, up to impact parameters of about 8 fm, there are
regions in the
Pb--Pb system where the density exceeds that in central S--U collisions. 
The transverse energy $dE_T/dy$ achieved in central collisions is roughly 
proportional to $N_p(0)$. From this one deduces that the average energy density
produced in central collisions, proportional to $N_p(0)/R^2$, is approximately 
the same in the S--U and Pb--Pb systems. 
However, the
maximum density achieved in Pb--Pb is about  35\% larger than in S--U. One 
may get an
estimate of the maximum value of $n_p$ by using sharp sphere densities. One
gets then
$n_p^{\max}(b)=2\rho_0\sqrt{(R_A+R_B)^2-b^2}$, which, for central
collisions, is proportional
to
$A^{1/3}+B^{1/3}$.  Note  that the
the $J/\Psi$ production, being proportional to $T_A\,T_B$, occurs
dominantly in the regions
of largest density.

Following \cite{BO88}, we now model the effect of quark-gluon plasma
formation by assuming that the
$J/\Psi$ produced  at point ${\bf s}$ is completely
destroyed whenever the density at that point exceeds a critical value. 
That is, we calculate 
\begin{equation}\label{Ndebplasma}
{\cal N}(b) =\frac{1}{T_{AB}(b)}\int {\rm d}^2s\,\frac{1}{\sigma_a^2}
\left( 1-{\rm e}^{-\sigma_aT_A({\bf s})} \right)
\left( 1-{\rm e}^{-\sigma_aT_B({\bf s-b})} \right) \theta({n_c-n_p(s)}).
\end{equation}
A plot of ${\cal N}(b)$ as a function of centrality, defined as 
$N_p(b)/N_p(b=0)$, is 
shown in Fig. 3, for various values of the critical density $n_c$.  
Given the fact that no suppression is observed in S--U collisions 
other than nuclear absorption,
the critical density has to be bigger than the
highest value attained in S--U collisions,
i.e. $3.3$ fm$^{-2}$ (see Fig. 2). Choosing this particular value for
 $n_c$, one  obtains
${\cal N}_{AB}= 0.28$. Therefore,
$r_\Psi=0.28/0.43=0.66$, to be compared with
the value $r_\Psi=0.72$ obtained by NA50. Furthermore,
for central collisions, we find
 $r_\Psi(0)=0.17/0.39=0.44$, to be compared with the value
$r_\Psi(0)\approx 0.50$ of NA50. Thus
the two main observations of NA50 can be accounted for quantitatively by this
simple picture.

Within the present model, 
the value $r_\Psi=0.66$ is to be looked at as the lowest possible 
since we have considered the most extreme scenario: total suppression above 
$n_c$ and lowest possible value of $n_c$.
The fact that the resulting $r_\Psi$ is only slightly smaller
than the experimental one puts very severe constraints on the
suppression mechanism. It seems for instance difficult to arrive at a
value of $r_\Psi$ as small as 0.72 by a mechanism which would set 
in gradually beyond S--U, and/or suppress only a fraction of the 
$\Psi$'s, as in a scenario advocating the suppression of resonances decaying 
into $J/\Psi$, such as $\chi$ for example. In order to attain the value 
$r_\Psi\approx 0.7$, {\it all\/} resonances have to be suppressed above $n_c$. 

Since the suppression mechanism is very sensitive to the local energy
density, the obtained
value of
$r_\Psi$ will be also sensitive to a number of factors. For example,  it is
sensitive  to the
parametrization of the nuclear density.  Thus, by taking another common
parametrization
($R_A=1.19A^{1/3}-1.61 A^{-1/3}$,
$a=0.54$~fm)  which makes the S nucleus smaller,
we get $r_\Psi=0.72$  instead of $0.66$. 
The deformation of the Uranium nucleus could also slightly alter the value 
of $r_\Psi$. We should also mention that in comparing
the two systems S--U and Pb--Pb, we have neglected the  variation of the
colliding energy, from 200~GeV to 160~GeV.  Such an energy shift results in a
slight decrease of the multiplicity density (in p--p collisions, this can be
estimated from \cite{ua5} to be about 4{\%}).  On the other hand, the energy
density is likely  to
increase with the number of participants faster than linearly, as was
assumed in our
calculation.  These effects should be taken into account in a more complete
calculation. It
should  also be stressed that none of the results on which we rely are
totally model independent, since 
the ratios $r_\Psi$ are obtained after extraction of nuclear absorption. 

It would obviously be  desirable to get 
confirmations of the present scenario from independent observables.  
In particular, it is worth recalling that the existence of a threshold 
effect is not a clear--cut theoretical prediction\cite{Matsui89}. 
It is therefore crucial to confirm experimentally such an effect  and
to determine the corresponding density; this could be achieved by exploring
collisions with smaller
targets. There are also several effects which could be looked for in the
Pb--Pb data,  and which,
if observed, would give confidence in the overall picture. These include
the effect of
fluctuations at large transverse energy, and the expected saturation with
centrality of the
 ratio
$r_{\Psi'}/r_\Psi$ and of the average transverse momentum squared
of the $J/\Psi$'s. We now discuss these three points.

It can be seen in Fig.3 that $r_\Psi$ is very sensitive
to the value of the critical density.
One can get a simple estimate  by considering
sharp sphere nuclear densities, and by neglecting nuclear absorption.
Then $r_\Psi (0)=(n_c/n_p^{\rm max})^4$ for $n_p^{\rm max}>n_c$,
in rough agreement with the results displayed in Fig.3.
We can write this same ratio in terms of energy densities:
$r_\Psi (0)=(\epsilon_c/\epsilon_{\rm max})^4$. Thus, $\epsilon_c$ being
fixed, a 10\% increase in $\epsilon_{\rm max}$ due to fluctuations,  leads to
a decrease of about 30\% in
$r_\Psi(0)$. One could therefore observe
a further noticeable decrease of $r_\Psi$ in collisions involving the
largest transverse energies.

It has been observed in the S--U system that the ratio $r_{\Psi'}/r_\Psi$
decreases with
increasing centrality. Because there are evidences that the $J/\Psi$ and
the $\Psi'$
suffer the same nuclear absorption \cite{NA38QM95}, it is natural to
attribute the extra
suppression to collisions with comovers (it  is plausible  that
the loosely bound
$\Psi'$ is more easily destroyed in hadronic collisions than the
$J/\Psi$).  However, in the present scenario, both the
$J/\Psi$ and the $\Psi'$ are destroyed before the comovers have a chance to
do anything. As a result,   the ratio
$r_{\Psi'}/r_\Psi$  remains approximately constant as a function of
centrality, as soon as the critical density is reached, i.e. for $b<b_c$.
This is easily deduced from Eq.(\ref{Ndebplasma}): when $b\le b_c$,
$r_{\Psi'}(b)/r_\Psi(b)\approx {\cal N}_{com}(b_c){\cal N}_{\Psi'}(b_c)/{\cal
N}_\Psi(b_c)$, and is approximately independent of $b$, while it 
should decrease if no plasma is produced.  
We have made a crude estimate, using the model discussed in \cite{BO88} 
of the quantity
${\cal N}_{com}$, which is the survival probability of the $J/\Psi$ after
its  interactions
with comovers. 
The values that we obtain  depend somewhat
 on parameters. However,
the saturation of the ratio
$r_{\Psi'}(b)/r_\Psi(b)$ with increasing centrality is a fairly robust
consequence of the
model.

 A natural explanation for
the  variations of the $J/\Psi$
$p_T$-distributions  observed in nuclear collisions  has been given in
terms of initial state
scatterings
\cite{pT}. In this picture, the increase of $\langle p_T^2\rangle$
at impact parameter $b$ is given by
$
\langle p_T^2\rangle=\langle p_T^2\rangle_0+C\,\bar n_{AB}(b)
$
where $C$ is a constant whose value can be determined from proton-nucleus
data, and
\begin{equation}\label{npABdeb}
\bar n_{AB}(b)=\frac{1}{ T_{AB}(b)}\int {\rm d}^2 s \,
T_A({\bf s})T_B({\bf s-b})
\left[T_A({\bf s})+T_B({\bf s-b})\right]\,{\cal N}(b)
\end{equation}
is the average density of nucleons seen by a $J/\Psi$.
The factor ${\cal N}(b)$ has been left out in previous analysis. 
However, it is important 
whenever the suppression is large. In particular, it is responsible here for
the fact that
$\bar n_{AB}(b)$ remains roughly constant when $b<b_c$, while in the absence 
of a plasma, $\bar n_{AB}(b)$ would continue to increase by some 25{\%}. 
This result, at variance with early expectations that a quark gluon plasma
would strongly affect the $J/\Psi$ momentum distribution\cite{BO90}, comes 
from the fact that in the present scenario {\it all\/} the $J/\Psi$'s need 
to be suppressed, irrespective of their transverse momentum, when $b<b_c$. 

In conclusion, we have explored a scenario in which 
$J/\Psi$ production is totally suppressed in regions
where the energy density  exceeds some critical value.
Quantitative agreement with the present 
 NA50 data is obtained by choosing the critical
density slightly greater than the density  attained in
central S--U collisions. The fact that the maximum
densities reached in the Pb--Pb and S--U systems may
differ by no more than 35\% suggests a strong sensitivity
of the suppression mechanism to small changes in the energy
density. It makes it  difficult to interpret the
present Pb--Pb data in terms of collisions with comovers
(such an interpretation is also made difficult by the
theoretical arguments developed in \cite{KS94}). 
It is therefore tempting to speculate that the large
increase in the suppression is due to a dramatic change in
the properties of the produced matter, pointing to the
possible production of the quark-gluon plasma. We wish to
stress however that the picture presented in
this letter is very crude, and although it appears to 
account for the bulk features of the present data, many
refinements need to be worked out, and
detailed confrontations with more data need to be made,
before unambiguous conclusions can be drawn.  

\medskip
\noindent {\bf Acknowledgments}

We wish to thank the members of the NA50 Collaboration for
discussions concerning their data.

\begin{figure}
\centerline{\epsfbox{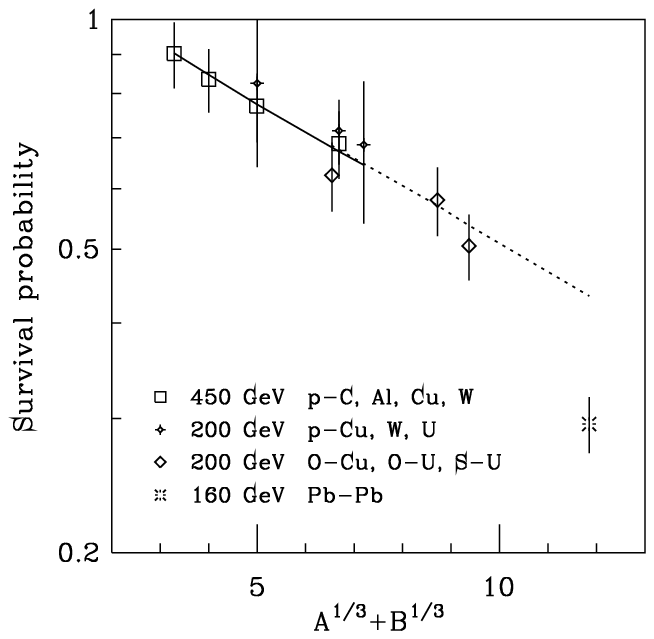}}
\caption{
\narrowtext
The $J/\Psi$ survival probability after absorption 
through nuclear matter, as a function of $A^{1/3}+B^{1/3}$, where $A$ and $B$ 
are the mass numbers of the colliding objects. The full line (dotted
line) is the survival probability for  the proton--nucleus (nucleus--nucleus)
systems, calculated with a cross section $\sigma_a=6.2$mb. 
The data at 450~GeV are obtained from refs.[1] and [8], 
those at 200~GeV from refs.[1] and [7].
In order to obtain the survival probability, each set of data has 
been rescaled by a constant factor so as to obtain the best fit 
to the theoretical curve.}
\end{figure}

\begin{figure}
\centerline{\epsfbox{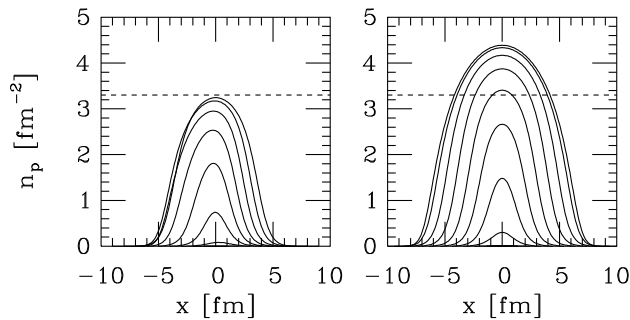}}
\caption{
The density of participants $n_p(s)$, for $s$ along the
direction of the impact parameter, for various values of the impact parameter: 
$b=0,2,4\cdots$~fm.
The origin is  at a distance $b/(1+R_B/R_A)$ from the center of nucleus A. left:
S--U collision; right: Pb--Pb collision. The horizontal dashed line
corresponds to the largest
density achieved in the S--U system, $n_p=3.3 {\rm fm}^{-2}$.
}
\end{figure}

\begin{figure}
\centerline{\epsfbox{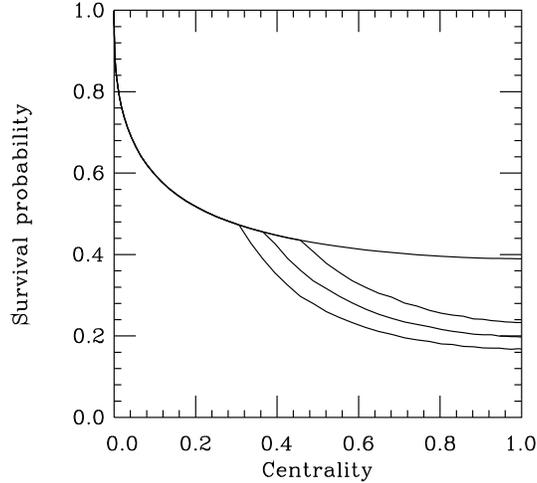}}
\caption{
The survival probability of a $J/\Psi$ in Pb--Pb
collisions after
absorption in nuclear matter and dissolution in a quark-gluon plasma
(eq.(\ref{Ndebplasma})).  For
values
$n_c >4.4$ fm$^{-2}$, there is no suppression beyond nuclear
absorption. The three
 curves showing an effect of the quark-gluon plasma correspond to $n_c$
=3.7, 3.5 and 3.3
respectively. The corresponding values of the ratio $r_\Psi$
(eq.(\ref{rpsi})) are respectively 0.82, 0.74 and 0.66.}
\end{figure}

\end{document}